\documentclass[prl,nobalancelastpage,twocolumn,nofootinbib,superscriptaddress,showpacs]{revtex4-1}
\usepackage{bm}
\usepackage{color}
\usepackage{amsfonts}
\usepackage{amssymb}
\usepackage{amsmath}
\usepackage{epsfig}
\usepackage{graphicx}
\usepackage{enumerate}
\usepackage{multirow}
\usepackage{verbatim}
\usepackage{color}
\usepackage{subfigure}

\newcommand{\ket}[1]{\vert#1\rangle}

\newcommand{\dg}{\dagger}

\newcommand{\mb}{\mathbf}

\begin{document}

\title{Chiral projected entangled-pair state with topological order}

\author{Shuo Yang}
\affiliation{Max-Planck Institut f\"ur Quantenoptik, Hans-Kopfermann-Str.~1, D-85748 Garching, Germany}
\affiliation{Perimeter Institute for Theoretical Physics, Waterloo, ON, N2L 2Y5, Canada}

\author{Thorsten B.~Wahl}
\affiliation{Max-Planck Institut f\"ur Quantenoptik, Hans-Kopfermann-Str.~1, D-85748 Garching, Germany}

\author{Hong-Hao Tu}
\affiliation{Max-Planck Institut f\"ur Quantenoptik, Hans-Kopfermann-Str.~1, D-85748 Garching, Germany}

\author{Norbert Schuch}
\affiliation{JARA Institute for Quantum Information, RWTH Aachen University, D-52056 Aachen, Germany}

\author{J.~Ignacio Cirac}
\affiliation{Max-Planck Institut f\"ur Quantenoptik, Hans-Kopfermann-Str.~1, D-85748 Garching, Germany}

\pacs{71.10.Hf, 73.43.-f}

\begin{abstract}
We show that projected entangled-pair states (PEPS) can describe chiral topologically ordered phases. For that, we construct a simple PEPS for spin-1/2 particles in a two-dimensional lattice. We reveal a symmetry in the local projector of the PEPS that gives rise to the global topological character. We also extract characteristic quantities of the edge conformal field theory using the bulk-boundary correspondence.
\end{abstract}

\maketitle

\textit{Introduction.---} Tensor network techniques provide a powerful tool to describe and analyze certain strongly correlated systems \cite{Ver08}. In particular, states corresponding to topologically ordered phases in lattices, like the Kitaev toric code \cite{Kit03}, resonating valence-bond states \cite{Ver06}, Levin-Wen string nets \cite{Lev05}, or their generalizations \cite{Bue08,Gu08} have very simple descriptions in terms of PEPS \cite{Ver04}, a particular family of states that can be described in terms of a simple tensor. Despite being a global property, the topological character of such states can be identified in terms of a symmetry group of the so-called auxiliary indices of such a tensor \cite{Sch10} (see also \cite{Bue14,Bur14}). The symmetry allows one to construct strings of operators that can be moved without changing the state, in terms of which one can identify the degeneracy of the parent Hamiltonian, the topological entropy \cite{Kit06b,Lev06}, or even the braiding properties of the excitations \cite{Sch10}. This, together with the bulk-boundary correspondence for PEPS \cite{Cir11}, as well as the existing numerical algorithms to determine correlation functions, the entanglement spectrum \cite{Hal08} and boundary Hamiltonians \cite{Cir11,Poi12,Sch13} provides us with a very powerful technique to analyze a variety of topologically ordered states (TOS).

The examples above do not contain any chiral TOS. Those states are utterly important, as they naturally appear in the fractional quantum Hall effect \cite{Sto99}, one of the most intriguing phenomena of modern physics. Furthermore, they can be associated to a topological quantum field theory, which dictate their coarse-grained properties, and relate them to other areas of mathematical and high energy physics. Whether PEPS (or other tensor network states) can describe examples of such states or not is a relevant question. An affirmative answer would open up the possibility of using the PEPS techniques to describe them, offering a new perspective to such important states and establishing a connection with other non-chiral TOS, like the ones mentioned above. A negative one would indicate that the family of PEPS should be extended in order to describe some of the most relevant strong-correlation phenomena.

A first hint that at least certain classes of TOS should be describable by simple PEPS was reported in Refs. \cite{Wah13,Dub13} (see also \cite{Ber11}). There, examples of fermionic states in lattices with non-trivial Chern numbers were reported. Those do not correspond to topologically ordered phases, as they are Gaussian and do not possess a connection with non-trivial topological quantum field theories. Nevertheless, these Gaussian fermionic PEPS (GFPEPS) share some of the properties of TOS like the existence of chiral edge modes, or the topological protection with respect to local perturbations provided by the Chern number \cite{Wah14}. The mere existence of the GFPEPS gives a strong expectation of the existence of other PEPS describing TOS. On the one hand, it is well known that some TOS can be constructed by applying the Gutzwiller projector technique on several copies of some particular Chern insulators or superconductors \cite{Wen89} (see also \cite{Par07,Gre09,Tu13}). On the other, it is rather obvious that this technique does not change the PEPS character of the states. Consequently, the states obtained by taking two or more copies of the states reported in Refs. \cite{Wah13,Wah14} and applying the Gutzwiller projector offer us a very promising candidate for a PEPS with TOS.

In this Letter, we analyze the state created by applying the Gutzwiller projector to two copies of chiral GFPEPS. The resulting state is a spin-1/2 PEPS with fermionic bonds; that is, where the auxiliary particles involved in the PEPS projectors are fermions \cite{Kra09,Cor10,Poi14}, with one (Dirac) fermion per bond. We develop methods to determine the boundary theory (including the boundary Hamiltonian and the corresponding entanglement spectrum) for such a PEPS defined on a cylinder, and to extract the corresponding conformal dimensions of the associated conformal field theory (CFT), which characterizes the topological order \cite{Wen90,Moo91}. The symmetries of the tensor and the boundary theory indicate that there are four primary fields, with conformal dimensions coinciding with those of the SO$(2)_1$ CFT. This, together with the value of the topological entropy, confirms that our state is indeed a TOS, thus providing the first example of a chiral PEPS with such a property.

The nature of the chiral GFPEPS used in the construction is akin to the $p+ip$ superconductor \cite{Rea00}, as it belongs to the same class according to the standard classification \cite{Sch08,Kit08}. This also explains why the state we obtain is associated to the SO$(2)_1$ CFT, since using the techniques of \cite{Tu13} it can be shown that this is the case for the state obtained by Gutzwiller projecting two copies of the $p+ip$ state. However, as opposed to that state, the chiral GFPEPS possesses correlations decaying as a power law with the distance, $r$ \cite{Wah13,Dub13}. This indicates that it corresponds to the state at the phase transition for any local parent Hamiltonian \cite{Wah14}. At the same time, there exists another parent Hamiltonian with long-range hoppings (scaling as $1/r^3$), which is gapped and which protects the chiral edges against perturbations \cite{Wah14}. By analyzing the gap in the transfer matrix \cite{Fan92}, we conclude that the PEPS we construct has infinite correlation length, and thus one can also view it as either at a topological phase transition or as the ground state of a gapped long-range Hamiltonian, which gives the topological robustness. While it is possible to determine a local parent Hamiltonian, we have not been able to identify the one with long-range interactions.

\textit{Spin-1/2 PEPS.---} Let us start with an $N_v \times N_h$ square lattice with periodic boundaries and spin-$\tfrac12$ \textit{physical} fermions at each site, with annihilation operator $a_{\mathbf{r},\alpha}$, where $\mathbf{r}$ denotes the lattice site and $\alpha=1,2$ the spin index. To construct the PEPS, we allocate eight additional \emph{virtual} Majorana modes at each site, denoted by $c^v_{\mathbf{r},\alpha}$ (with $v=L,R,U,D$
and $\alpha=1,2$). Below we suppress the site index $\mathbf{r}$ when considering a single site, for ease of reading. To each site, we associate a fiducial state $\ket{P_{\bm r}}\equiv\ket{P}$ (corresponding to the ``PEPS projector'') which we choose site-independent in order to ensure translation invariance. It can be written as an entangled state out of physical and virtual fermionic modes
\begin{equation}
|P\rangle=P |\Omega \rangle =\left(a_1^{\dag}b_1^{\dag}+a_2^{\dag}b_2^{\dag}\right)|\Omega
\rangle,  \label{eq:projector}
\end{equation}
where $b_{\alpha}=\frac{1}{2\sqrt{2}}[(c^L_{\alpha}+ic^R_{\alpha})e^{i \tfrac{\pi}{4}} -(c^U_{\alpha}-ic^D_{\alpha})]$ is an annihilation operator acting on the virtual modes, and the vacuum $|\Omega\rangle$ is annihilated by $d_{\alpha}=\frac{1}{2\sqrt{2}} [(c^L_{\alpha}+ic^R_{\alpha})e^{i \tfrac{\pi}{4}}+(c^U_{\alpha}-ic^D_{\alpha})]$ in addition to $a_{\alpha}$ and $b_{\alpha}$. To complete the PEPS construction, we project the virtual
Majorana modes between neighboring sites onto maximally entangled Majorana bonds $\omega_{\mathbf{r},\mathbf{r}+\mathbf{x}}= \prod_{\alpha=1}^{2}\tfrac{1}{2}(1+ ic^R_{\mathbf{r},\alpha} c^L_{\mathbf{r}+\mathbf{x},\alpha})$ and
$\omega_{\mathbf{r},\mathbf{r}+\mathbf{y}}= \prod_{\alpha=1}^{2} \tfrac{1}{2}(1+ ic^D_{\mathbf{r},\alpha} c^U_{\mathbf{r}+\mathbf{y},\alpha})$, see Fig.~\ref{fig:PEPS}~\cite{Kra09}. After discarding the virtual modes, we arrive at the PEPS wave function
\begin{align}
|\Psi\rangle = \langle \Omega^{\prime}| \prod_{\langle \mathbf{r},\mathbf{r}^{\prime} \rangle} \omega_{\mathbf{r},\mathbf{r}^{\prime}} \prod_\mathbf{r} P_\mathbf{r} |\Omega \rangle, 
\label{eq:PEPS}
\end{align}
where $|\Omega^{\prime} \rangle$ indicates the vacuum of the auxiliary fermions.

\begin{figure}[tbp]
\includegraphics[width=\columnwidth]{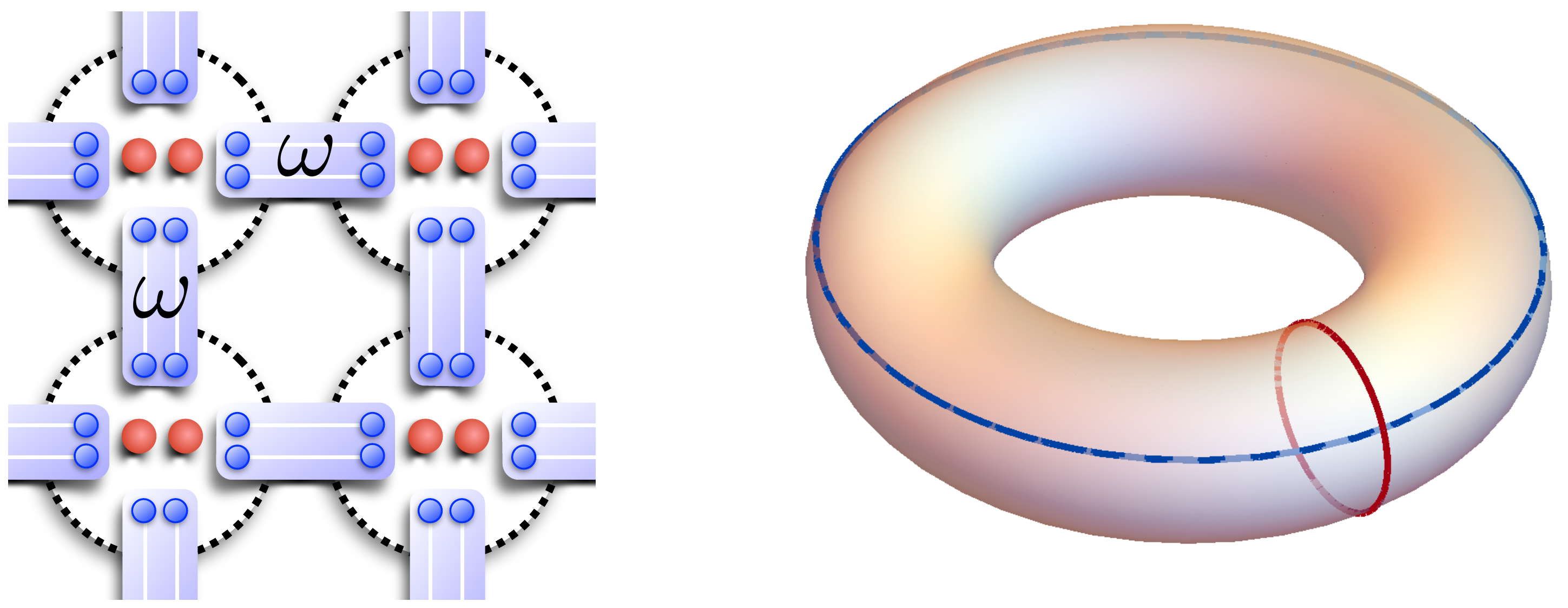}
\caption{Left: Construction of the PEPS defined via
Eqs.~\eqref{eq:projector} and~\eqref{eq:PEPS}.  $|P\rangle$ is defined on
two physical fermionic modes (red) and eight virtual Majorana modes (blue)
for each site. Sites are marked by dashed circles.
Afterwards, the projectors $\omega_{\mb r, \mb r + \mb x}$ and
$\omega_{\mb r, \mb r + \mb y}$ are applied on the four Majorana modes
located between neighboring sites, as indicated by light blue boxes,
yielding the PEPS $|\Psi\rangle$.
Right: Non-contractible loops on the torus.}
\label{fig:PEPS}
\end{figure}

The state (\ref{eq:PEPS}) describes spin-1/2 particles in a lattice, as there is always a single fermion per site. Furthermore, it is not Gaussian, as $\ket P$ is itself not Gaussian, and thus it must correspond to a theory with interactions. In fact, $\ket P$ can be viewed as a Gaussian state $|\Phi \rangle =\prod_{\alpha =1}^{2}(\sqrt{1-\lambda }+\sqrt{\lambda }a_{\alpha }^{\dag
}b_{\alpha }^{\dag })|\Omega \rangle $ (with $0<\lambda <1$), followed by a Gutzwiller projection $P_{\mathrm{G}}=a_{1}^{\dagger}a_{1}a_{2}a_{2}^{\dagger }+a_{1}a_{1}^{\dagger }a_{2}^{\dagger }a_{2}$ enforcing \emph{single} occupancy of physical fermions, $|P\rangle =P_{\mathrm{G}}|\Phi \rangle $. Without $P_{\mathrm{G}}$, it is known that $|\Psi \rangle $ is a product of two identical GFPEPS, which individually describe a topological superconductor of spinless fermions \cite{Wah14}. The long-range parent Hamiltonian of this GFPEPS has Chern number $C=-1$ and supports a single chiral Majorana edge mode at each boundary, implying that it belongs to the same class as the $p+ip$ topological superconductor \cite{Rea00}.

The close relation between the GFPEPS used in our construction and the $p+ip$ state suggests that $|\Psi\rangle$ could have the same associated CFT as the wave function constructed by Gutzwiller projecting two copies of the $p+ip$ state. As argued in \cite{Tu13}, the latter is a chiral topological state with Abelian anyons, whose edge theory is an SO(2)$_{1}$ [or equivalently, U(1)$_{4}$] CFT. Such a CFT describes a chiral Luttinger liquid with central charge $c=1$ and four primary fields $I$, $s$, $\bar{s}$, and $v$, whose conformal dimensions are $h_{I}=0$, $h_{s}=h_{\bar{s}}=1/8$, and $h_{v}=1/2$, respectively. This serves as a guide for our interacting PEPS construction and provides sharp predictions which we can compare with numerics. Of course, one has to keep in mind that, unlike the $p+ip$ state, the GFPEPS has powerlaw decaying quasi-long-range correlations, so we do not have \emph{a priori} knowledge whether topological order is still present in the PEPS (\ref{eq:PEPS}).

\begin{figure*}[tbp]
\centering
\includegraphics[width=0.95\linewidth]{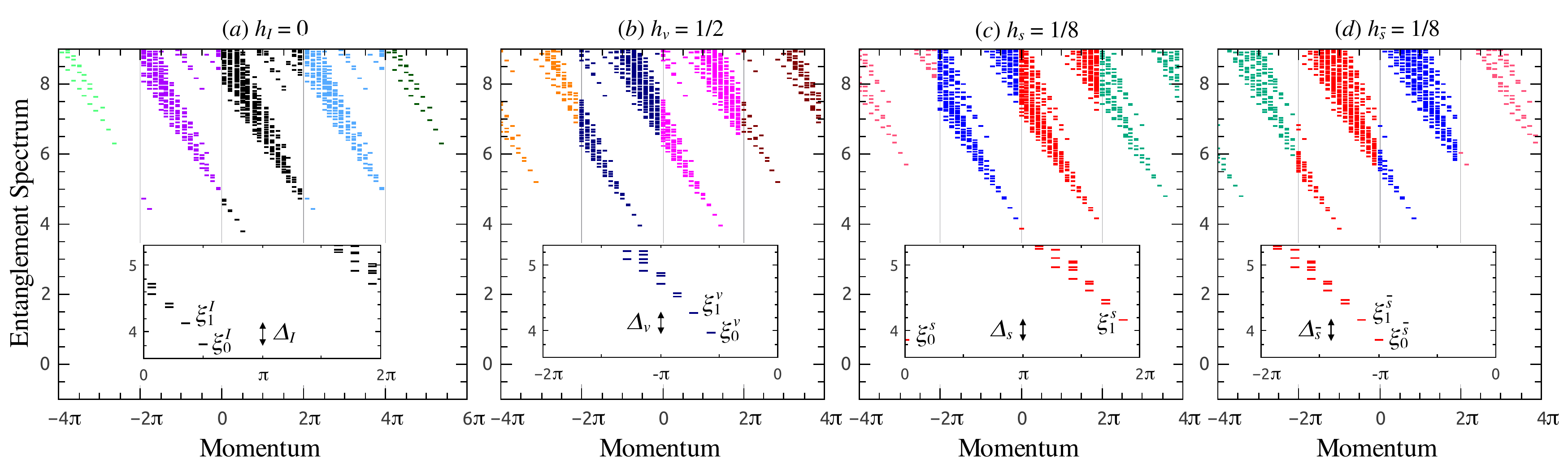}
\caption{(Color online) The entanglement spectra for a bipartition of a $N_v=14$ infinite cylinder in four topological sectors. The momentum in horizontal direction is extended beyond its $2\protect\pi$ periodicity to distinguish the spectra corresponding to different particle-number subspaces. The entanglement spectra with the same particle number obey the degeneracy pattern $\left\{1,1,2,3,5,\cdots \right\}$, a fingerprint for $c=1 $ chiral Luttinger liquid theory. Insets: Zoom-in of the low-lying part of the spectra.}
\label{fig:ES}
\end{figure*}

\textit{Symmetry.---} The symmetry of the local PEPS description is known to be important for topological states. For $\ket P$ in (\ref{eq:projector}), we find three $\mathbb{Z}_2$ gauge symmetries
\begin{align}
U_0 |P\rangle := X_{L}X_{R}X_{U}X_{D}|P\rangle &= -|P\rangle ,  \label{eq:flux} \\
U_{\alpha} |P\rangle &= |P\rangle , \label{eq:symmetry}
\end{align}%
where $\alpha = 1,2$, $X_{v}=ic_{1}^{v}c_{2}^{v}$ ($v=L,R,U,D$) and $U_{\alpha} = {\openone} - 2d_\alpha^\dg d_\alpha$. $X_{v}$ and $U_\alpha$ are unitaries and $X_{v}^{2}=U_\alpha^2 = {\openone}$. In fact, $U_0$ measures the fermion parity of the virtual modes, which arises due to the single-occupancy constraint of physical fermions and is absent without the Gutzwiller projection $P_{\mathrm{G}}$. The symmetry $U_{\alpha}$ is inherited from the free fermionic state $|\Phi\rangle$~\cite{Wah14}.

A bosonic version of $U_0$ appears in the toric code as a $\mathbb{Z}_2$-injectivity~\cite{Sch10}, whose generalizations allow to explain the properties of all known non-chiral topological phases~\cite{Sch10,Bue14,Bur14}. A similar formalism allows us to also uncover the topological character from the symmetry in the present scenario. For instance, a \textit{closed} loop of operators $\prod_{v\in C}X_{v}$, where the loop $C$ crosses the virtual bonds, can be inserted between the virtual bonds and the PEPS projectors in (\ref{eq:PEPS}). If $C$ is contractible, the symmetry condition $U_0$ allows us to remove the loop operator. In contrast, when the PEPS is defined on a manifold with nontrivial topology, \textit{non-contractible} loops $C$ exist and nontrivial loop operators $W(C)=\prod_{v\in C}X_{v}$ can be defined (see Fig.~\ref{fig:PEPS}). The symmetry condition $U_0$ implies that such loop operators can be moved around in the PEPS without affecting the physical state.  In particular, states both with and without loops are locally described by the same ``bare'' PEPS. Thus, they are all ground states of local \emph{parent Hamiltonians} $H=\sum_{j}h_{j}$, where the $h_j$ enforce the PEPS structure locally (i.e., $h_j$ is local, $h_j\ge0$, and $\mathrm{ker}\, h_j$ is given by all states which look locally like the PEPS \cite{Supp}).

Instead of inserting loop operators defined by the symmetry $U_0$ (called ``fluxes''), it is also possible to insert loop operators constructed from $U_\alpha$ (called ``strings''), which can likewise be deformed without changing the state~\cite{Wah14} as long as they do not cross $U_0$-type loops. By combining these possibilities, five different ground states of the local parent Hamiltonian can be constructed~\cite{Supp}.  Four of them (characterized by $U_0$-type loops) are topological, while the fifth arises from the fact that the local parent Hamiltonian is gapless, i.e., has a continuous spectrum above the ground state energy.

\textit{Boundary theory.---} In order to characterize the topological nature of the state, we compute the entanglement spectrum of the minimally entangled states (MES)~\cite{Zha12}.  We consider a bipartition of a torus into cylinders: There, the four MES are characterized by the presence or absence of a $U_0$ flux \emph{along} the cylinder, and are eigenstates of a $U_0$ loop \emph{around} the cylinder~\cite{Sch13,Supp}.  In the context of PEPS, the entanglement spectrum (this is, the spectrum of the reduced density operator of the cylinder) can be determined from the reduced state of the \emph{virtual} fermions at the boundaries of the cylinder~\cite{Cir11}. For the MES, the two boundaries of the cylinder decouple in the limit of a long cylinder, and the entanglement spectrum for e.g.\ the left boundary is given by $\sqrt{\sigma_L^\top} \sigma_R \sqrt{\sigma_L^\top} \propto \varrho_L =:e^{-H_L}$~\cite{Cir11}, where $\sigma_{L}$ ($\sigma_R$) is the reduced density matrix for the virtual modes on the left (right) boundary, $\varrho_L$ is normalized to $\mathrm{Tr} \varrho_L=1$, and we have implicitly defined the entanglement Hamiltonian $H_L$. In Ref.~\cite{Supp} we show how $\sigma_L$ and $\sigma_R$ can be determined numerically for each symmetry sector.

\begin{figure}[tbp]
\centering
\includegraphics[width=0.95\columnwidth]{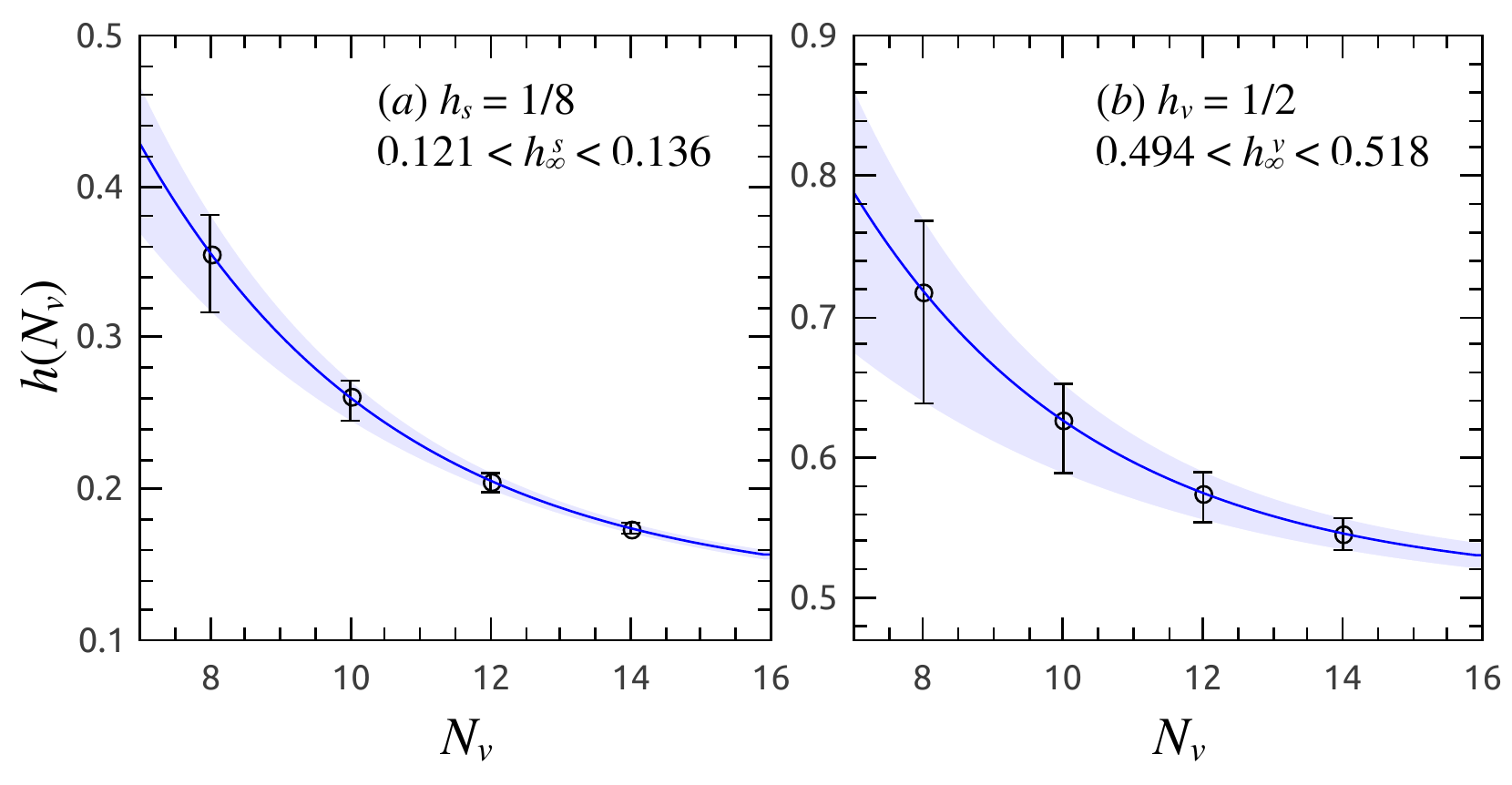}
\caption{
\label{fig:spin}
(Color online) Conformal dimensions of the primary fields (a) $s$
and (b) $v$ through exponential decay fitting.}
\end{figure}

From the spectrum of the entanglement Hamiltonian, we are able to extract the conformal dimensions of the CFT primary fields, based on the theory developed in Ref. \cite{Qi12}: For each MES $|\Psi _{\mu }\rangle $, the reduced density matrix of a cylinder cut from a torus is a thermal state of two \emph{chiral} CFTs restricted to the sector $\mu $ (labeling the primary field generating the tower of states), $\varrho _{\mu }\propto e^{-H_{L}-H_{R}}|_{\mu }$, where $H_{L}$ and $H_{R}$ live at the left and right boundaries of the cylinder, respectively. When constraining to the left boundary, the spectrum of the PEPS boundary Hamiltonian for the MES $|\Psi _{\mu}\rangle $, at least its low-energy part, should correspond to the chiral CFT spectrum of $H_{L}$ restricted to the sector $\mu $. Once this is settled, the procedure of extracting the conformal dimensions of primary fields is the following: (i) The sector with the lowest entanglement energy $\xi _{0}^{I}$ corresponds to the CFT identity field $I$ with $h_{I}=0$; (ii) The differences of the two lowest entanglement energies $\xi _{0}^{\mu }$ and $\xi _{1}^{\mu }$ in the same sector $\mu $, $\Delta _{\mu }=\xi_{1}^{\mu }-\xi _{0}^{\mu }$, set the energy scale of the CFT. This energy scale is the same for all sectors, $\Delta _{\mu }=\Delta $; (iii) The difference of lowest level entanglement energies in sectors $\mu $ and $I$, divided by the energy scale $\Delta $, gives the conformal dimension of primary field $\mu $, $h_{\mu }=(\xi _{0}^{\mu }-\xi _{0}^{I})/\Delta$.

The entanglement spectra of the four MES $|\Psi _{\mu }\rangle $ with $N_{v}=14$ and $N_{h}=\infty$ are shown in Fig.~\ref{fig:ES}. In all four sectors, the state counting of the spectra shows clear chiral Luttinger liquid behavior with the characteristic degeneracy pattern $\{1,1,2,3,5,\ldots\}$. Following the described procedure, we extract the conformal dimension $h_{\mu }$ for each sector, shown in Fig.~\ref{fig:spin}. Since the $\Delta _{\mu }$ are slightly different due to finite-size effects we use their average, and indicate the minimal and maximal values by error bars. To obtain an estimate of the conformal dimensions in the thermodynamic limit,  we use a fit $h(N_{v})=h_{\infty}+A\exp (-N_{v}/t)$ and find that $h_{\text{avg},\infty }^{s}=0.131$, $0.121<h_{\infty }^{s}<0.136$ (same for $\bar{s}$), $h_{\text{avg},\infty }^{v}=0.510$, and $0.494<h_{\infty }^{v}<0.518$, in good agreement with the SO(2)$_{1}$ CFT prediction.

\begin{figure}[tbp]
\centering
\includegraphics[width=0.95\columnwidth]{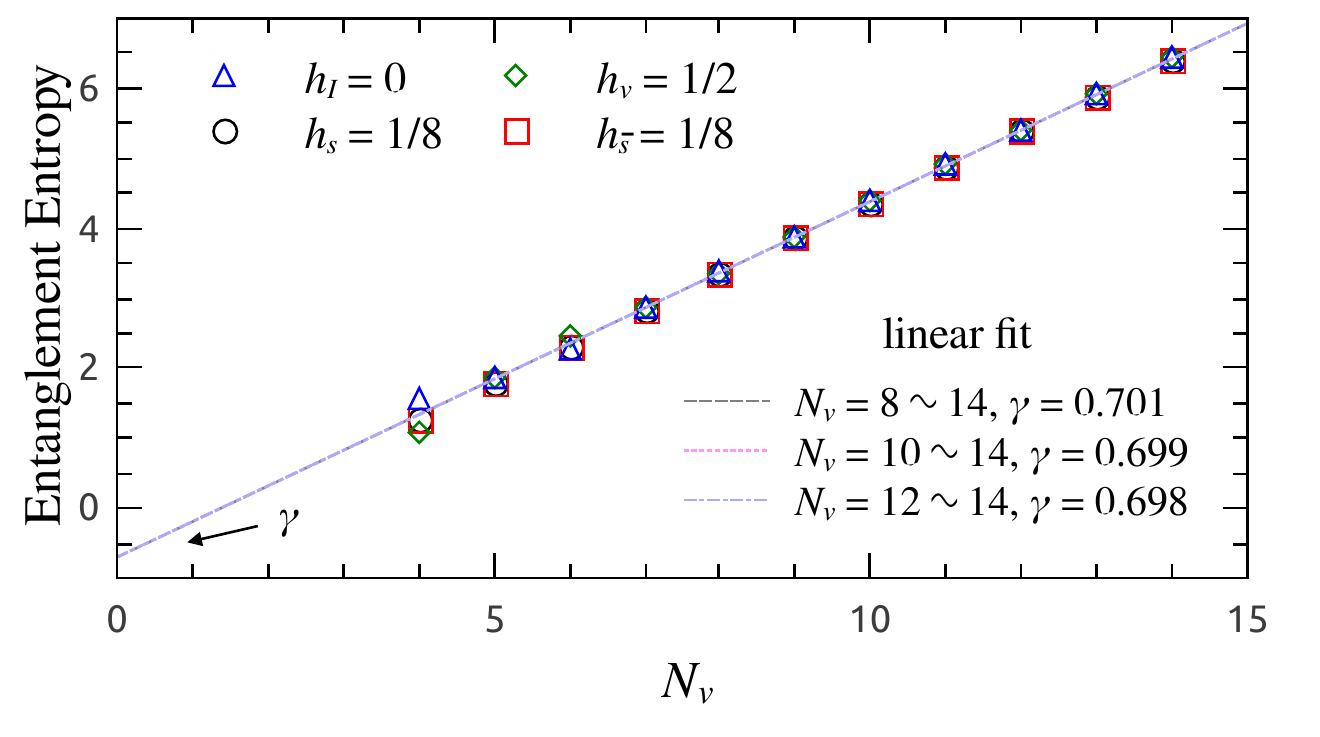}
\caption{(Color online) Entanglement entropy versus perimeter $N_{v}$ for
different topological sectors. The dashed lines are linear fits based on the
average of the four data points for $N_{v}=8\sim 14$, $N_{v}=10\sim 14$, and
$N_{v}=12\sim 14$, respectively.}
\label{fig:TEE}
\end{figure}

The spectra of the boundary Hamiltonian for the MES also give direct access to the von Neumann entropy $S_{\mathrm{vN}}$ of the reduced density matrix on a half-infinite cylinder. For sectors with anyonic excitations, the von Neumann entropy of the MES $|\Psi _{\mu }\rangle $ contains the usual area law contribution and a universal subleading constant $\gamma_{\mu}=\ln(\mathcal{D}/d_{\mu})$~\cite{Zha12}, where $d_{\mu }$ is the quantum dimension of the anyonic quasiparticle $\mu$ and $\mathcal{D}$ the total quantum dimension, $\mathcal{D}=\sqrt{\sum_{\mu }(d_{\mu })^{2}}$. For the $\mathrm{SO(2)}_{1}$ theory, there exist only four Abelian anyons with $d_{\mu }=1$, and thus $\mathcal D=2$.  Fig.~\ref{fig:TEE} shows the von Neumann entropies of the MES of the example (\ref{eq:PEPS}) as a function of $N_{v}$. We find that the difference between the sectors vanishes as $N_v$ increases, and fitting $S_{\mathrm{vN}}(N_{v})=\alpha N_{v}-\gamma_\mu $ \cite{Jia12} gives $\gamma_\mu\approx\ln 2$, in agreement with the prediction of $\mathrm{SO(2)}_{1}$ theory.

\textit{Transfer operator.---} Let us now address whether our interacting chiral PEPS has infinite correlation length, as it has been found for the GFPEPS describing topological superconductors and insulators \cite{Wah13,Dub13}. In the PEPS formalism, this is related to the absence of a gap of the transfer operator in the limit $N_v \rightarrow \infty$. Using the numerical technique sketched in the Supplemental Material \cite{Supp}, we have determined such a gap for different values of $N_v$ (both with and without flux) for the PEPS (\ref{eq:PEPS}). The results are shown in Fig.~\ref{fig:Gap}, and suggest that the gap of the transfer operator vanishes polynomially in the thermodynamic limit, indicating a divergent correlation length. The same conclusion can be drawn by studying the interaction range of the boundary Hamiltonian \cite{Supp}.

\begin{figure}[tbp]
\centering
\includegraphics[width=0.95\columnwidth]{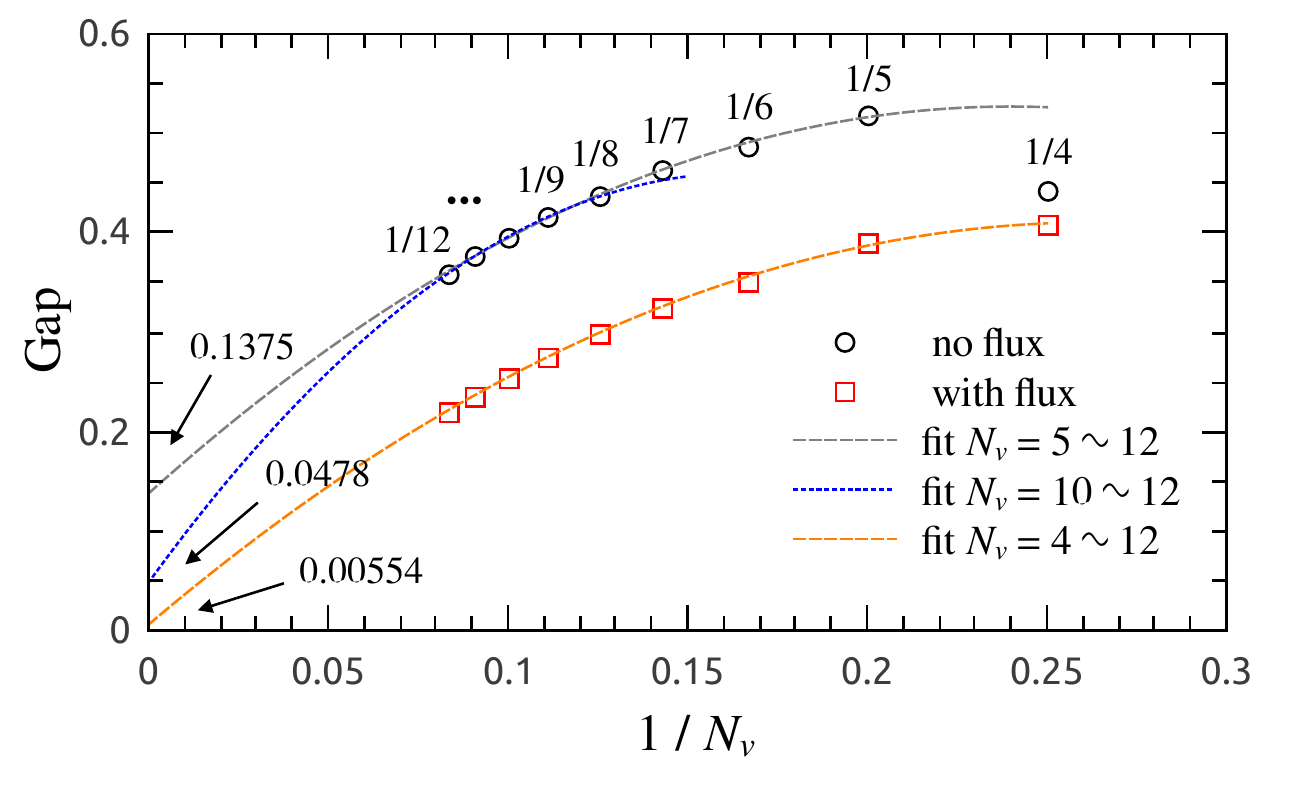}
\caption{ \label{fig:Gap}
(Color online) Finite-size scaling of the gap of the transfer
operators with and without flux. The fitting function is
$a/N_{v}^{2}+b/N_{v}+c$, with $c$ given in the plot (for the ``no flux''
sectors, fits for two different ranges of $N_v$ are given), and suggests
that the gap vanishes as $1/N_v$ in the thermodynamic limit.}
\end{figure}

\textit{Conclusions.---} In this Letter, we have investigated to which extent PEPS can describe chiral topological order, by constructing and studying an explicit example. We have identified the local symmetries in the PEPS, known to be responsible for the topological characters of all non-chiral topological states.  Based on these symmetries, we further showed that our example exhibits the characteristic properties of topologically ordered chiral states, such as ground state degeneracy, an entanglement spectrum described by a chiral CFT, and nonvanishing topological entanglement entropy. Finally, we have provided numerical evidence suggesting that the state has a diverging correlation length and therefore a gapless local parent Hamiltonian. This raises several intriguing questions: Can one construct an interacting chiral topological PEPS which has exponentially decaying correlations and is the ground state of a gapped local Hamiltonian? Can our PEPS describe the state in a topological phase transition? And, is it possible to determine a parent Hamiltonian with long-range interactions which stabilizes the topological phase?

\textit{Acknowledgments.---} Parts of this work were done at the Perimeter Institute for Theoretical Physics (Waterloo), the Simons Institute for the Theory of Computing (Berkeley), the Centro de Ciencias de Benasque Pedro Pascual (Benasque), and the Erwin Schr\"{o}dinger International Institute for Mathematical Physics (Vienna). HHT thanks X.-L.~Qi and J.~Dubail for helpful discussions. Part of this work has been supported by the EU projects SIQS and QALGO, and the Alexander von Humboldt foundation. Research at Perimeter Institute is supported by the Government of Canada through Industry Canada and by the Province of Ontario through the Ministry of Economic Development \& Innovation.


\onecolumngrid
\appendix
\setcounter{equation}{0}
\newpage

\renewcommand{\thesection}{S-\arabic{section}} \renewcommand{\theequation}{S%
\arabic{equation}} \setcounter{equation}{0} \renewcommand{\thefigure}{S%
\arabic{figure}} \setcounter{figure}{0}

\centerline{\textbf{Supplemental Material}}

\maketitle

\section{S-1. Symmetries of the PEPS}

In the following we explain how to obtain the five states on the torus, which are the ground states of a local Hamiltonian, which we also show how to be constructed. We then argue that only four correspond to the topological sectors, while the fifth arises from the fact that for the local Hamiltonian we have a gapless continuum of excitations.

\subsection{A. Construction of the five ground states on the torus}

We will consider states which are made out of the original PEPS by
inserting non-contractible loops of operators constructed from the
symmetries
\begin{align}
U_0 |P\rangle = X_{L}X_{R}X_{U}X_{D}|P\rangle &=-|P\rangle ,  \label{eq:flux}
\\
U_{\alpha} |P\rangle &= |P\rangle ,  \label{eq:string}
\end{align}
($\alpha = 1,2$). As outlined in the main paper, Eq.~\eqref{eq:flux} gives
rise to loops $C$ of the form $\prod_{v \in C} X_v$, which can be moved
without changing the state; we will call these loops \emph{fluxes}. They are called fluxes because they correspond to threading a $\pi$ flux through the torus, which gives rise to a phase of $\pi$ when moving an electron around the torus.
Similarly, Eq.~\eqref{eq:string} gives rise to loops of string operators
(called \emph{strings} in the following) of the form $\sum_{v \in C}
c^v_\alpha$, which can be moved freely as well~\cite{S_Wah14}. Due to the
construction of parent Hamiltonians as operators which enforce that the
state is locally described by the PEPS, any PEPS with such movable strings will
still be a ground state. By combining such loops in horizontal and vertical
direction, we can in principle construct $2^{2 \cdot 3} = 64$ states on the
torus; we will denote these states by $|s_{1h}, s_{1v}, s_{2h}, s_{2v}, f_h,
f_v\rangle$ (string in first layer along horizontal/vertical direction,
string in second layer along horizontal/vertical direction, flux along
horizontal/vertical direction; e.g., $s_{1h} = 1 (0)$ denotes the presence
(absence) of a string along the horizontal direction in the first layer).

Yet, as we will see in the following, this only gives rise to five linearly
independent ground states. This stems from three facts: First, certain
states vanish (i.e., have norm zero); second, states can be linearly
dependent; and third, different loops might not commute with each other,
which makes it impossible to move their crossing point. 

Let us first consider the state defined on a single layer $\alpha = 1, 2$
prior to the Gutzwiller projection: As shown in~\cite{S_Wah14}, if one
contracts the PEPS on a horizontal cylinder with open virtual indices on its
ends, the state of physical and virtual modes has a virtual fermionic mode
in the vacuum state delocalized between the two edges of the cylinder at $%
k_y = 0$ (we are allowed to consider the Fourier transform of the modes in
vertical direction, since the state is a Gaussian fermionic state). However, the
projection on maximally entangled Majorana modes when closing the horizontal
boundary corresponds to a projection on occupied fermionic modes delocalized
between the two edges of the cylinder. Thus, the state on the torus
vanishes. In contrast, the insertion of one or two strings occupies the
fermionic mode at $k_y = 0$ (for a horizontal string it occupies the one at $%
k_x = 0$ of a vertical cylinder), rendering the final state non-vanishing.
Note that the state is the same for an insertion of a horizontal or a
vertical string and has a different parity as the state obtained for two
strings. Those arguments are also valid after the application of the
Gutzwiller projector, as it only acts on the physical modes.

Let us now turn to the state after the Gutzwiller projection, and consider
the case without any fluxes. We are left with four different possibilities,
as on each layer either one or two strings can be inserted. Since the
Gutzwiller projector keeps only states with total parity even, that is, with
the same parity on both layers, there are only two different states without
fluxes, which are $|101000\rangle \propto |100100\rangle \propto
|011000\rangle \propto |010100\rangle$ and $|111100\rangle$.

Next, let us consider a state with one flux along the vertical direction.
One can easily verify that a flux crossing a string cannot be moved freely
(it induces a phase jump of $\pi$ in the string), raising its energy and
thus ruling out horizontal strings. We are thus left with the possibility of
inserting vertical strings in either layer. But if one does insert any of
these, the final state is vanishing: The insertion of a string occupies the
fermionic mode at $k_y = 0$ on the corresponding layer, as pointed out
above, while the flux inverts the final projection on occupied modes to a
projection on empty modes. The same argument of course applies to a
horizontal flux, leaving us with two non-vanishing states $|000010\rangle$
and $|000001\rangle$.

Finally, a state with two fluxes does no longer allow for the insertion of
strings, since the fluxes could no longer be moved freely, and thus yields $%
|000011\rangle$ as the last ground state.
We have verified numerically that the five remaining states are indeed
linearly independent on a $4 \times 4$ and a $5 \times 4$ torus. 

\subsection{B. Parent Hamiltonian}

The parent Hamiltonian can be constructed using standard PEPS techniques \cite{S_Ver08}. For that, we have constructed a $4 \times 4$ plaquette, and found a maximal operator, $h=h^2 \ge 0$, that annihilates the PEPS on that plaquette for all values of the anxiliary indices (i.e., leaving them unprojected). The parent Hamiltonian is obtained by adding all horizontal and vertical translations of $h$.

Starting from the operator $h$ acting on a $4 \times 4$ plaquette, and adding all possible translations, we have constructed a Hamiltonian for the $4 \times 4$ and $5 \times 4$ lattice. We have verified that the ground space of that Hamiltonian is five-fold degenerate, with the eigenstates exactly corresponding to the ones constructed in the previous subsection.

\subsection{C. Topological sectors}

Let us argue that four ground states correspond to distinct
topological sectors, while the fifth is the lowest state in a gapless
continuum of excitations. To this end, consider the construction of a $%
\mathrm{SO}(2)_{1}$ theory by Gutzwiller projecting two $p+ip$ states \cite%
{S_Tu13}. These $p+ip$ states are ground states of the pairing Hamiltonian%
\begin{equation}
H_{p+ip}=\sum_{\mathbf{k}\alpha }[2(\cos k_{x}+\cos k_{y})-\mu ]a_{\mathbf{k}%
\alpha }^{\dagger }a_{\mathbf{k}\alpha }+\Delta \sum_{\mathbf{k}\alpha
}(\sin k_{x}+i\sin k_{y})(a_{\mathbf{k}\alpha }^{\dagger }a_{-\mathbf{k}%
\alpha }^{\dagger }+a_{-\mathbf{k}\alpha }a_{\mathbf{k}\alpha }),
\label{eq:p+ip}
\end{equation}%
where $\alpha $ denotes the two layers ($\alpha =1,2$) and we choose $0<\mu
<4$ and $\Delta \neq 0$ so that the model is in the topological phase. The $%
p+ip$ Hamiltonian (\ref{eq:p+ip}) can have periodic or anti-periodic
boundary conditions in either direction, but the boundary conditions must be
the same in each layer to ensure translational invariance. All four states
have even fermion parity and thus survive after Gutzwiller projection. These
Gutzwiller projected states are four topologically distinct states
corresponding to the quasiparticle types of the $\mathrm{SO}(2)_{1}$ theory.
On the other hand, in the PEPS, the choice of anti-periodic boundary
conditions corresponds to the insertion of a flux loop, leading us to the
conclusion that the four topological ground states are distinguished by
their flux loop configuration.

It remains to understand why there are two states in the flux-free sector.
To this end, consider a single layer of the gapless topological
superconductor used to construct the PEPS. Its parent Hamiltonian has a
gapless band of excitations with minimum at $\mathbf{k}=(0,0)$. On the other
hand, anti-periodic boundary conditions shift the momentum by $\pi /N$,
i.e., only in the flux-free sector the gapless band will give rise to a
second ground state. This carries over through the Gutzwiller projection,
since it enforces equal parity for both copies of the superconductor.
Following this reasoning and Ref.~\cite{S_Wah14}, we find the state with an
empty $\mathbf{k}=(0,0)$ mode, $|101000\rangle $, to be the ground state,
while $|111100\rangle $ arises from the gapless continuum above it. Note
that this is also compatible with the construction of a Gutzwiller projected
state from the fully gapped $p+ip$ Hamiltonian (\ref{eq:p+ip}) with periodic
boundary conditions in both directions.

\section{S-2. Numerical implementation}

In the following, we give a description of the numerical method used to
construct the boundary density operator for our chiral PEPS. We restrict ourselves to an $N_{v} \times N_{h}$ cylinder with periodic boundary conditions along the
vertical ($v$) direction [Fig.~\ref{appendix:cylinder}]. Finally we take the limit of infinite cylinders with $N_h \rightarrow \infty$.

\subsection{A. Contraction of GFPEPS}

For the
double-layer PEPS introduced in the main text, we define in each lattice
site a plaquette containing two physical fermionic modes and four virtual
fermionic modes. The annihilation operators of the virtual modes are denoted
as $L$ (left), $R$ (right), $U$ (up), and $D$ (down), which are constructed
by combining the Majorana modes of the first layer and second layer in the
same direction. That is $v=\left(c_{1}^{v}-i c_{2}^{v}\right)/2$ for $v=L$, $R$, $U$
and $D$.

\begin{figure}[tbp]
\includegraphics[width=0.4\columnwidth]{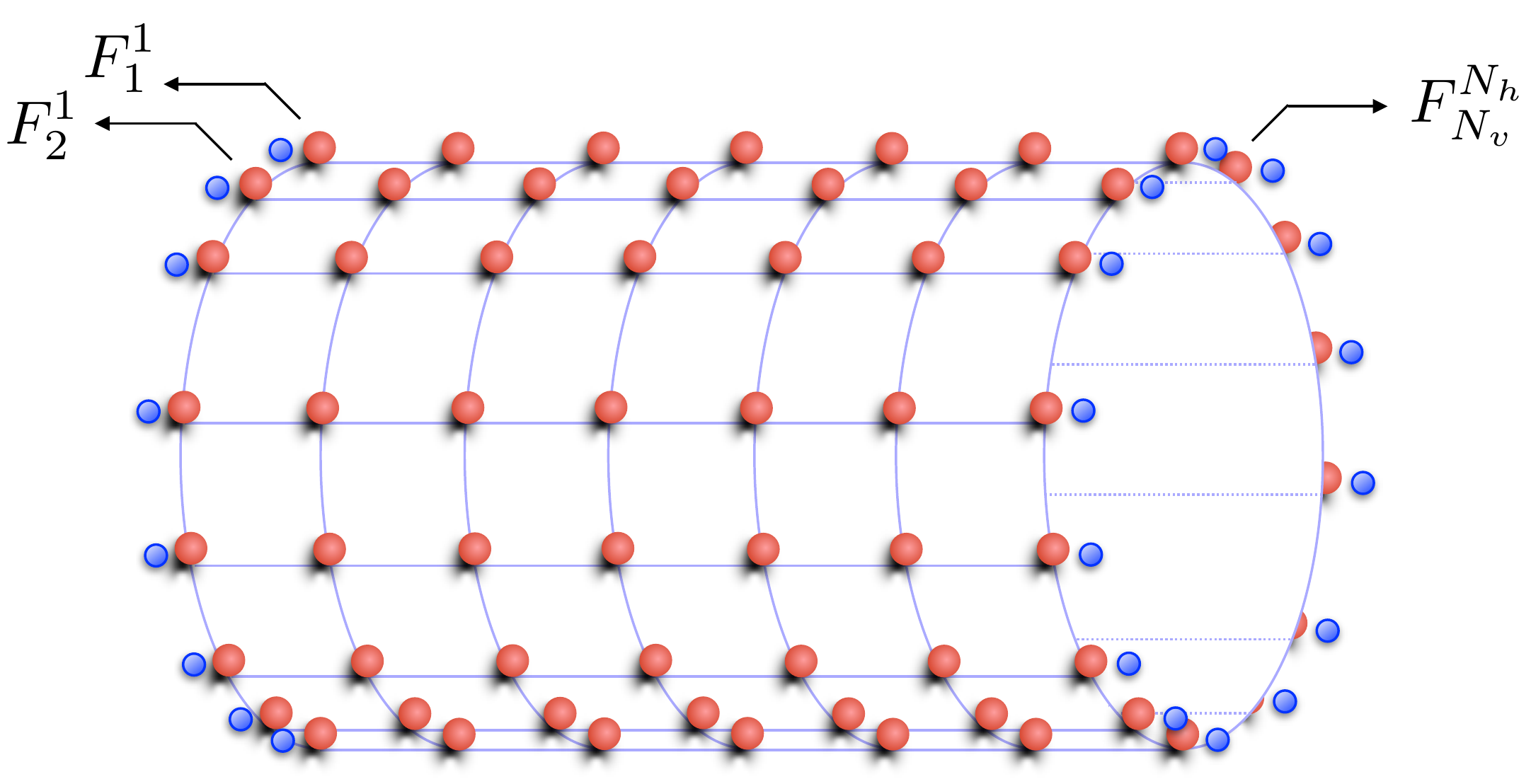}
\caption{(Color online) Construction of the PEPS on an $N_{v} \times N_{h}$
cylinder.}
\label{appendix:cylinder}
\end{figure}

We write the fiducial state as $\left| \Psi \right\rangle=F\left| \Omega
\right\rangle$, where $F$ is an operator made out of the creation and
annihilation operators acting on a plaquette, and $\left| \Omega
\right\rangle$ is the vacuum. We label the rows from up to down and the
columns from left to right. Each plaquette $F_{i}^{j}$ is labelled by its
row $i$ and column $j$ [Fig.~\ref{appendix:cylinder}]. We will omit the
indices whenever there is no ambiguity. Since we have periodic boundary
conditions along the vertical direction, the row index is understood as mod $%
N_v$.

We denote $\omega_{i,i+1}^{j}$ as the maximally entangled operator acting on
two neighboring virtual fermions $D_{i}^{j}$ and $U_{i+1}^{j}$,
\begin{align}
\omega &=(\openone+ic_{1}^{D} c_{1}^{U})(\openone+ic_{2}^{D} c_{2}^{U})/4  \notag \\
&=(D^{\dagger} D U U^{\dagger}+D D^{\dagger} U^{\dagger} U+i D^{\dagger} U
+i D U^{\dagger})/2.
\label{omega}
\end{align}
Similarly, we denote $\eta_{i}^{j,j+1}$ as the maximally entangled operator
acting on $R_{i}^{j}$ and $L_{i}^{j+1}$.
\begin{align}
\eta =(R^{\dagger} R L L^{\dagger}+R R^{\dagger} L^{\dagger} L +i
R^{\dagger} L +i R L^{\dagger})/2.
\end{align}
We note that
\begin{align}
\omega^{2}=\omega =\omega^{\dagger},  \notag \\
\eta^{2}=\eta =\eta^{\dagger}.  \label{omegaeta}
\end{align}
Also, all the $\omega$'s and $\eta$'s commute among themselves, since they
act on different fermionic modes and are even in the number of fermionic
mode operators.

We denote by $F^{j}=\prod_{i} F_{i}^{j}$, $\omega^{j}=\prod_{i}
\omega_{i,i+1}^{j}$, and $\eta^{j,j+1}=\prod_{i} \eta_{i}^{j,j+1}$. With all
that, we define the wave function
\begin{align}
\left| \Phi_{N_{h}} \right\rangle=\left[ \prod_{j=1}^{N_h-1} \eta^{j,j+1} %
\right] \left[ \prod_{j=1}^{N_{h}} \omega^{j} \right] \left[
\prod_{j=1}^{N_{h}} F^{j} \right] \left| \Omega \right\rangle.
\end{align}
If we ignore all the modes that are projected by the $\omega$'s and $\eta$%
's, this is the state for: (i) $N_{v} \times N_{h}$ physical modes $%
p_{i}^{j} $; (ii) $N_{v}$ virtual modes living at the left of the cylinder $%
L_{i}^{1}$; (iii) $N_{v}$ virtual modes living at the right, $R_{i}^{N_{h}}$
[Fig.~\ref{appendix:cylinder}].

We are interested in the boundary theory of the right virtual modes. That is
the density operator $\sigma_{R}$ acting on the modes $R_{i}^{N_{h}}$ alone.
In order to obtain that, we will specify an initial density operator $x^{0}$
in which we project the modes $L_{i}^{1}$. The operator $\sigma_{R}$ is
defined such that for any operator $G$ acting on the $R_{i}^{N_{h}}$,
\begin{align}
\text{tr}\left( \sigma_{R} G \right)=\left\langle \Phi_{N_{h}} \right| G
x^{0} \left| \Phi_{N_{h}} \right\rangle.
\end{align}
To write this expression in a more suitable form, we define
\begin{align}
x^{n} &=\left\langle \Omega^{n} \right| (F^{n})^{\dagger}
(\omega^{n})^{\dagger} (\eta^{n,n+1})^{\dagger} x^{n-1} \eta^{n,n+1}
\omega^{n} F^{n} \left| \Omega^{n} \right\rangle,  \notag \\
&=\text{tr} \left[ x^{n-1} \eta^{n,n+1} \omega^{n} F^{n} \left| \Omega^{n}
\right\rangle \left\langle \Omega^{n} \right| \left( F^{n} \right)^{\dagger} %
\right]  \label{makerow}
\end{align}
where $\Omega^{n}$ represents the vacuum for all the modes in column $n$. We
have used that certain operators commute and Eq.~(\ref{omegaeta}). Since $%
\omega$ does not depend on the $R$'s and commutes with $x^{N_{h}-1}$, we can
now write
\begin{align}
\sigma_{R}=\text{tr}\left[ x^{N_{h}-1} \omega^{N_{h}} F^{N_{h}} \left|
\Omega^{N_{h}} \right\rangle \left\langle \Omega^{N_{h}} \right|
(F^{N_{h}})^{\dagger} \right].  \label{makelastrow}
\end{align}
The trace is with respect to all the operators in the last row except for the $R$%
's. Therefore, for an input $x_{0}$, we will determine $x^{n}$ successively
using Eq.~(\ref{makerow}), and in the end obtain $\sigma_{R}$ with Eq.~(\ref%
{makelastrow}).

To be specific, we start out with $x^{n-1}$ and add plaquettes one by one to
obtain $x^{n}$ [Eq.~(\ref{makerow})]. Adding plaquette $F_{1}^{n}$ gives
\begin{align}
y_{1}^{n} &=\mathrm{tr}_{L_{1}^{n}} \left[ x^{n-1} G_{1} \right],
\end{align}
where $G_{1} =\mathrm{tr}_{p_{1}^{n},U_{1}^{n},D_{1}^{n},R_{1}^{n}} $ $[
\eta_{1}^{n,n+1} \omega_{N_{v},1}^{n} \omega_{1,2}^{n} $ $F_{1}^{n} $ $%
\left| \Omega_{1}^{n} \right\rangle $ $\left\langle \Omega_{1}^{n} \right| $
$(F_{1}^{n})^{\dagger} ]$. This is an operator acting on $L_{1}^{n}$, $%
U_{2}^{n}$, $D_{N_{v}}^{n}$, and $L_{1}^{n+1}$. We can do the same for $2
\leq m \leq N_{v}-1$,
\begin{align}
y_{m}^{n} =\mathrm{tr}_{U_{m}^{n},L_{m}^{n}} \left[ y_{m-1}^{n} G_{m}\right].
\end{align}
Here $G_{m} =\mathrm{tr}_{p_{m}^{n},D_{m}^{n},R_{m}^{n}} $ $[
\eta_{m}^{n,n+1} \omega_{m,m+1}^{n} $ $F_{m}^{n} \left| \Omega_{m}^{n}
\right\rangle $ $\left\langle \Omega_{m}^{n} \right| $ $(F_{m}^{n})^{%
\dagger} ] $. For the last plaquette in row $n$, we have
\begin{align}
x^{n} =\mathrm{tr}_{U_{N_{v}}^{n},D_{N_{v}}^{n},L_{N_{v}}^{n}} \left[
y_{N_{v}-1}^{n} G_{N_{v}} \right],
\end{align}
with $G_{N_{v}} =\mathrm{tr}_{p_{N_{v}}^{n},R_{N_{v}}^{n}} $ $[
\eta_{N_{v}}^{n,n+1} F_{N_{v}}^{n} \left| \Omega_{N_{v}}^{n} \right\rangle $
$\left\langle \Omega_{N_{v}}^{n} \right| (F_{N_{v}}^{n})^{\dagger} ]$. We
note that $G_{2}= \cdots = G_{N_{v}-1}$ up to relabeling of operators, and
that $G_{1}$, $G_{2}$, and $G_{N_{v}}$ are the same for all columns, so that
they have to be calculated only once.

Once we have obtained $x^{N_{h}-1}$, the plaquettes from the last row are
added one by one to get $\sigma_{R}$ [Eq.~(\ref{makelastrow})]. Adding
plaquette $F_{1}^{N_{h}}$ gives
\begin{align}
z_{1} =\mathrm{tr}_{L_{1}^{N_{h}}} \left[ x^{N_{h}-1} H_{1} \right],
\end{align}
where $H_{1} =\mathrm{tr}_{p_{1}^{N_{h}},U_{1}^{N_{h}},D_{1}^{N_{h}}} $ $[
\omega_{N_{v},1}^{N_{h}} \omega_{1,2}^{N_{h}} F_{1}^{N_{h}} $ $|
\Omega_{1}^{N_{h}} \rangle $ $\langle \Omega_{1}^{N_{h}} | $ $%
(F_{1}^{N_{h}})^{\dagger} ]$. Adding plaquette $F_{m}^{N_{h}}$ with $2 \leq
m \leq N_{v}-1$ gives
\begin{align}
z_{m} &=\mathrm{tr}_{U_{m}^{N_{h}},L_{m}^{N_{h}}} \left[ z_{m-1} H_{m} %
\right],
\end{align}
where $H_{m} =\mathrm{tr}_{p_{m}^{N_{h}},D_{m}^{N_{h}}} $ $[
\omega_{m,m+1}^{N_{h}} F_{m}^{N_{h}} $ $| \Omega_{m}^{N_{h}} \rangle $ $%
\langle \Omega_{m}^{N_{h}} | $ $(F_{m}^{N_{h}})^{\dagger} ]$. Finally, by
adding the last plaquette $F_{N_{v}}^{N_{h}}$ we have
\begin{align}
\sigma_{R}=\mathrm{tr}%
_{U_{N_{v}}^{N_{h}},D_{N_{v}}^{N_{h}},L_{N_{v}}^{N_{h}}} \left[ z_{N_{v}-1}
H_{N_{v}} \right],
\end{align}
where $H_{N_{v}}=\mathrm{tr}_{p_{N_{v}}^{N_{h}}} $ $[ F_{N_{v}}^{N_{h}} |
\Omega_{N_{v}}^{N_{h}} \rangle $ $\langle \Omega_{N_{v}}^{N_{h}} | $ $%
(F_{N_{v}}^{N_{h}})^{\dagger} ]$. Again, we just have to calculate $H_{1}$, $%
H_{2}$, and $H_{N_{v}}$, since $H_{m}=H_{2}$ up to relabeling for $%
m=3,\cdots,N_{v}-1$.

It is important to remark that one has to be extremely careful with the
definition of the trace and the vacuum when one deals with fermions. The
natural Hilbert space for the fermionic modes does not possess a tensor
product structure, so that one cannot simply write $\left| \Omega
\right\rangle=\left| \Omega^{1} \right\rangle \otimes \cdots \otimes \left|
\Omega^{N_{h}} \right\rangle$. Similarly, the trace is defined in terms of
an orthonormal basis, which will be built in terms of creation operators. It
cannot be moved inside an expression since those operators do not commute
with each other. The appropriate way of doing that is using a Jordan-Wigner
transformation (JWT), so that we can deal with spins. In particular, when
calculating the trace, we define the JWT such that the corresponding
operators are in the right order. We ensure that the operators we do not
trace do not have any strings corresponding to the ones we do trace. Then we
can trace the corresponding spins as usual. Alternatively, one may
explicitly calculate the signs caused by fermion anti-commutation relations
and absorb them into local tensors. In this way, the memory and CPU
requirements are largely reduced. Finally, the complexity of the calculation remains the same as for the spin models in Ref.~\cite{S_Cir11,S_Sch13,S_Poi12,S_Yan14}.

\subsection{B. Construction of minimally entangled states}

In this subsection, we show the fixed point density matrix $\sigma_L$ and $\sigma_R$ of each topological sector are determined numerically by (i) inserting or not inserting the $\mathbb{Z}_2$ flux operator through the cylinder, and (ii) imposing even- or odd-parity boundary conditions at the ends of the cylinder.

The insertion of a flux is equivalent to changing one row of maximally
entangled operators $\omega$ (e.g. $\omega_{1,2}^{j}$ for $j=1,2,\cdots,N_h$) from Eq.~(\ref{omega}) to
\begin{align}
\omega &=(\openone-ic_{1}^{D} c_{1}^{U})(\openone-ic_{2}^{D} c_{2}^{U})/4  \notag \\
&=(D^{\dagger} D U U^{\dagger}+D D^{\dagger} U^{\dagger} U-i D^{\dagger} U
-i D U^{\dagger})/2,
\end{align}
while all the other $\omega$ and all $\eta$ remain the same. 
Imposing even (odd) boundary conditions at the ends means that the eigenvalues of the virtual fermionic modes $\sum v^{\dagger}v$ for $v=L$ and $R$ are even (odd) for both bra and ket layers. 

We focus on the case with an even number of sites in the circumference direction hereafter.
In the presence of a flux, we obtain $\sigma_{I}$ when imposing even boundaries for $N_v=4m$ systems or imposing odd boundaries for $N_v=4m+2$ systems, where $m$ is an integer. Meanwhile, we obtain $\sigma_{v}$ when imposing odd boundaries for $N_v=4m$ systems or imposing even boundaries for $N_v=4m+2$ systems. 
In the absence of a flux, the fixed point reduced density matrix (RDM) $\sigma$ is in general a linear superposition of two minimally entangled states (MES) corresponding to topological sectors $s$ and $\overline{s}$. We may recover the true MES by minimizing the rank of RDM numerically. When imposing even-parity boundary conditions on both the bra and ket layers at both ends, we obtain the MES $\sigma_{se}$ and $\sigma_{\overline{s}e}$. (We refer to them as $\sigma_{s}$ and $\sigma_{\overline{s}}$ in the main text.) When imposing odd-parity boundary conditions, we obtain MES $\sigma_{so}$ and $\sigma_{\overline{s}o}$ in the same topological sectors $s$ and $\overline{s}$, respectively.

Once we obtain the fixed point RDM $\sigma_L$ and $\sigma_R$, the boundary RDM $\varrho_L$ which reproduces the entanglement spectrum is given by $\varrho_L \propto \sqrt{\sigma_L^{\top}} \sigma_R \sqrt{\sigma_L^{\top}}$ with $\mathrm{Tr}\varrho_L=1$. For the current system we find $\sigma_L^{\top}=\sigma_R$, so that $\varrho_L \propto \sigma_R^{2}$.

\section{S-3. Interaction range}

In the following, we explore the locality of the boundary Hamiltonians for the interacting system and compare the result with its non-interacting counterpart. 

\begin{figure}[tbp]
\centering
\includegraphics[width=0.65\columnwidth]{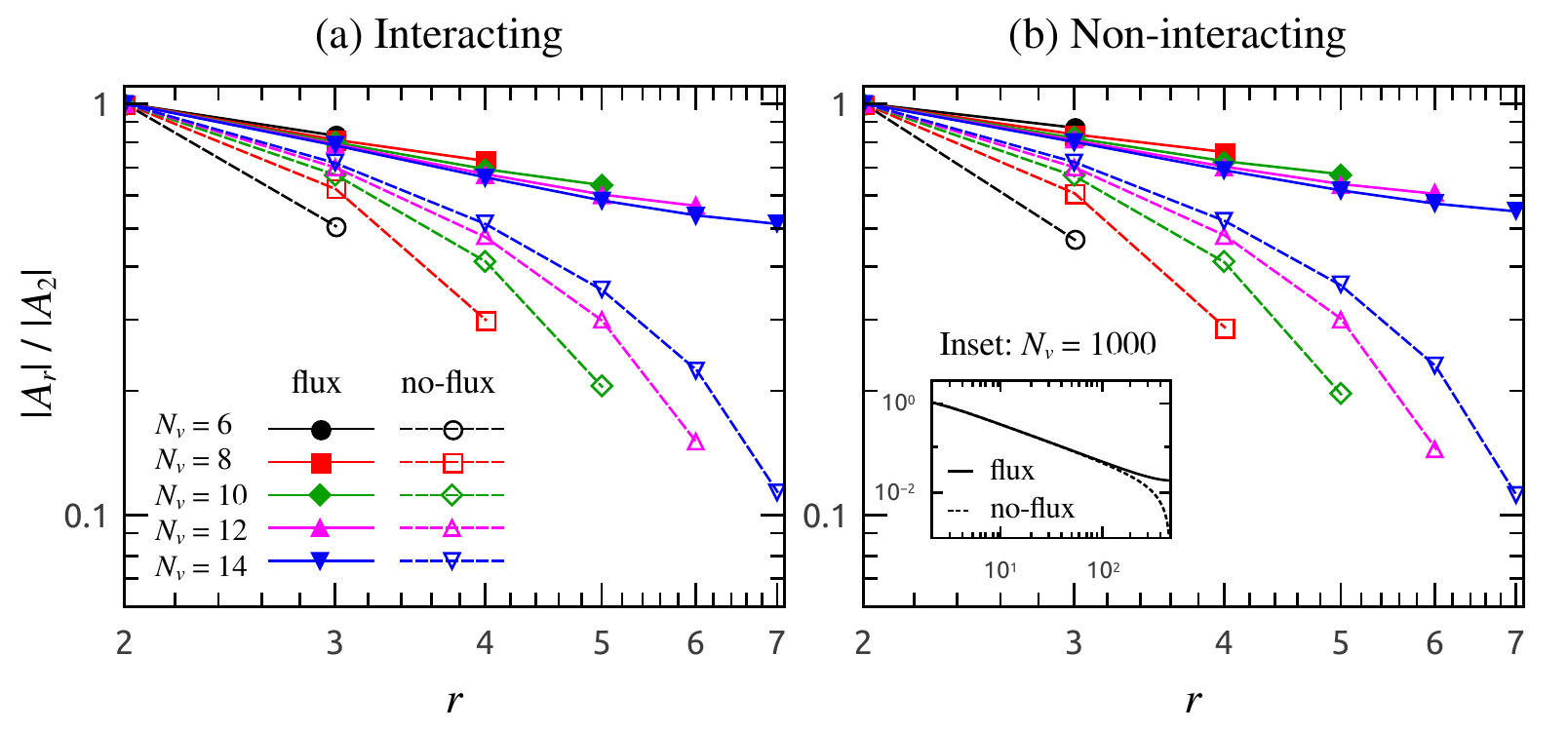}
\caption{(Color online) Interaction range of the boundary Hamiltonian with
and without flux, plotted using a log-log scale. (a) Interacting case. (b)
Non-interacting case. Inset: Non-interacting case with $N_v=1000$.}
\label{fig:IntRange}
\end{figure}

From the previous subsection we find the boundary RDM has the following form
\begin{align}
\varrho=w_{I}\varrho_{I}+ w_{v}\varrho_{v}+ w_{se}\varrho_{se}+ w_{\overline{s}e}\varrho_{\overline{s}e}+ w_{so}\varrho_{so}+ w_{\overline{s}o}\varrho_{\overline{s}o}.
\end{align}
where the weights $w$ are determined at the ends of the cylinder and the presence or absence of a flux. The individual $\varrho_{\{\cdots\}}$'s are normalized to $\mathrm{Tr} \varrho_{\{\cdots\}}=1$. 
To make the boundary Hamiltonians as local as possible, we choose the equal-weight combinations of RDM in the flux sector and no-flux sector \cite{S_Sch13}. We define two Hamiltonians $H_{\textrm{flux}}$ and $H_{\textrm{no-flux}}$,
\begin{align}
H_{\textrm{flux}}&=-\log \left[ (\varrho_{I}+\varrho_{v})/2 \right], \notag \\
H_{\textrm{no-flux}}&=-\log \left[ (\varrho_{se}+\varrho_{\overline{s}e}+\varrho_{so}+\varrho_{\overline{s}o})/4 \right].
\end{align}
Then we decompose the boundary Hamiltonian in terms of Majorana fermions:
\begin{align}
H=\sum_{i_{1},\cdots,i_{N_v}=0}^{3}f_{i_{1},\cdots,i_{N}} c^{i_{1}} c^{i_{2}} \cdots c^{i_{N_v}}.
\end{align} 
Here $c^{0}=I$, $c^{1}=c_{1}$, $c^{2}=c_{2}$, $c^{3}=c_{1}c_{2}$, and $c_{1}$ and $c_{2}$ are the virtual Majorana modes of the double copies. 
We say the locality of a term is $r$ if it spans $r$ nearest neighbor sites (taking the periodic boundary conditions into consideration) \cite{S_Cir11}. Thus $r=1$ denotes the constant term, $r=2$ denotes the nearest neighbor terms, $r=3$ includes both next-nearest neighbor terms and the terms acting on three contiguous sites, \textit{etc}. The interaction strength $|A_{r}|$ is defined as the $2$-norm of all the weights of terms with locality $r$ \cite
{S_Cir11,S_Sch13,S_Poi12,S_Yan14}.

In Fig.~\ref{fig:IntRange}(a) we show the relative interaction strength $|A_{r}|/|A_{2}|$ as a function of distance $r$. We find that the interactions in $H_{\textrm{flux}}$ and $H_{\textrm{no-flux}}$ indeed decay with distance. Moreover, the two curves of $H_{\textrm{flux}}$ and $H_{\textrm{no-flux}}$ will converge with increasing $N_v$. Although the finite size effect is strong, we may get some hints by comparing it with the non-interacting case \cite{S_Wah14}. In Fig.~\ref{fig:IntRange}(b), the non-interacting case shows almost the same behavior for $N_v=6,8,\cdots,14$. For the large-size systems, since the decay obeys a power law for the non-interacting case (see the inset of Fig.~\ref{fig:IntRange}(b) and Ref.~\cite{S_Wah14}), we would expect a power-law decay for the interacting case as well. This is consistent with the infinite correlation length indicated in the main text.

\end{document}